\documentclass[letter]{aa}

\usepackage{graphicx}
\usepackage{txfonts}
\usepackage{natbib}

\begin{document} 

\title{Terrestrial deuterium-to-hydrogen ratio in water in hyperactive comets}

\author{Dariusz C. Lis\inst{1,2}, Dominique Bockel\'{e}e-Morvan\inst{3}, Rolf G\"{u}sten\inst{4}, Nicolas Biver\inst{3}, J\"{u}rgen Stutzki\inst{5}, Yan Delorme\inst{2}, Carlos~Dur\'{a}n\inst{4}, Helmut Wiesemeyer\inst{4}, Yoko Okada\inst{5}}

\institute{Jet Propulsion Laboratory, California Institute of Technology, 4800 Oak Drove Drive, Pasadena, CA 91109, USA
  \and Sorbonne Universit\'{e}, Observatoire de Paris, Universit\'{e} PSL, CNRS, LERMA, F-75005, Paris, France
  \and LESIA, Observatoire de Paris, Universit\'{e} PSL, CNRS, Sorbonne Universit\'{e}, Universit\'{e} Paris Diderot, Sorbonne Paris Cit\'{e}, 5 place Jules Janssen, 92195 Meudon, France
  \and Max-Planck-Institut für Radioastronomie, Auf dem H\"{u}gel 69, D-53121 Bonn, Germany
  \and I. Physikalisches Institut, Universität zu K\"{o}ln, Z\"{u}lpicher Stra\ss e 77, D-50937 K\"{o}ln, Germany}

\date{$\copyright 2019$. All rights reserved. Received March 26, 2019; accepted April 11, 2019.}

\abstract{ The D/H ratio in cometary water has been shown to vary between 1 and 3 times the Earth’s oceans value, in both Oort cloud comets and Jupiter-family comets originating from the Kuiper belt. This has been taken as evidence that comets contributed a relatively small fraction of the terrestrial water. We present new sensitive spectroscopic observations of water isotopologues in the Jupiter-family comet 46P/Wirtanen carried out using the GREAT spectrometer aboard the Stratospheric Observatory for Infrared Astronomy (SOFIA). The derived D/H ratio of $(1.61 \pm 0.65) \times 10^{-4}$ is the same as in the Earth’s oceans.  Although the statistics are limited, we show that interesting trends are already becoming apparent in the existing data. A clear anti-correlation is seen between the D/H ratio and the active fraction, defined as the ratio of the active surface area to the total nucleus surface. Comets with an active fraction above 0.5 typically have D/H ratios in water consistent with the terrestrial value. These hyperactive comets, such as 46P/Wirtanen, require an additional source of water vapor in their coma, explained by the presence of subliming icy grains expelled from the nucleus. The observed correlation may suggest that hyperactive comets belong to a population of ice-rich objects that formed just outside the snow line, or in the outermost regions of the solar nebula, from water thermally reprocessed in the inner disk that was transported outward during the early disk evolution.  The observed anti-correlation between the active fraction and the nucleus size seems to argue against the first interpretation, as planetesimals near the snow line are expected to undergo rapid growth. Alternatively, isotopic properties of water outgassed from the nucleus and icy grains may be different due to fractionation effects at sublimation. In this case, all comets may share the same Earth-like D/H ratio in water, with profound implications for the early solar system and the origin of Earth’s oceans.}
 

\titlerunning{D/H ratio in hyperactive comets}
\authorrunning{Lis et al.}
\maketitle

\section{Introduction}

One of the key questions for modern astrophysics and planetary science concerns the development of the conditions of habitability in planetary systems, such as the early protosolar nebula. Water, an essential ingredient for carbon-based life as we know it \citep{Ref2}, is formed primarily via surface reactions in icy mantles of interstellar dust grains (the gas-phase chemistry only becomes efficient at temperatures above $\sim 300$~K). These grains subsequently find their way through dense protostellar cores to protoplanetary disks, where they are partially processed thermally in the warm inner disk before being locked up in small bodies such as comets or asteroids \citep{Ref3}.

In the standard model of the protosolar nebula, the temperature in the terrestrial planet forming zone was too high for water ice to survive.
Consequently, the Earth accreted dry and the present-day water would have been delivered in a later phase, together with organics, by external sources such as comets or asteroids \citep{Ref4}. An alternative explanation is in situ, early delivery of Earth's water, either incorporated into olivine grains or through the oxidation of an early hydrogen atmosphere by FeO in the terrestrial magma ocean, both of which may have contributed to some degree  (see \citealt{Ref4} and references therein).

The D/H ratio provides key constraints on the origin and thermal history of water molecules. Deuterium was produced in the Big Bang, with an abundance of about $2.5 \times 10^{-5}$ with respect to hydrogen \citep{Ref5}. The reference protosolar D/H ratio in hydrogen is $2.1 \times 10^{-5}$, which is close to the Big Bang value \citep{Ref6}. However, in the cold, dense, CO-depleted interstellar medium, deuterium atoms are preferentially sequestered in heavy molecules due to differences in zero-point vibrational energies \citep{Ref7}. Consequently, the D/H ratio in heavy molecules may be enhanced by orders of magnitude, and doubly or even triply deuterated species have been detected \citep{Ref8,Ref9}. Deuteration in water is less extreme than in other molecules, with water D/H ratios of order 0.001—0.01 typically measured in low-mass protostars similar to our Sun \citep{Ref7}.
Subsequent isotopic exchanges between water molecules and molecular hydrogen in the warm inner disk  drives the ratio back toward the protosolar value \citep{Ref10}.

The highest solar system D/H ratios in water, about $7.3 \times 10^{-4}$ measured in LL3 matrix clays or R chondrites \citep{Ref11,Ref12,Ref13}, are  close to the interstellar medium values. The D/H ratio in Earth’s ocean water, the Vienna Standard Mean Ocean Water (VSMOW), is significantly lower, $(1.5576 \pm 0.0001) \times 10^{-4}$, although still enhanced with respect to the protosolar ratio in hydrogen. How representative this value is for the bulk of Earth’s water is a subject of discussion in the light of recent measurements of a low D/H ratio in deep mantle materials \citep{Ref14}. Currently, carbonaceous chondrites, in particular CI and CM types, appear to best match the terrestrial D/H ratio \citep{Ref12}.

Comets are the most primitive volatile-rich bodies in the solar system. The D/H ratio has been measured in a handful of Oort cloud comets, with typical values of about twice VSMOW \citep{Ref1}. The \emph{Herschel} Space Observatory provided the first measurements of the D/H ratio in two Jupiter-family  comets, 103P/Hartley \citep{Ref15} and 45P/Honda-Mrkos-Pajdu\v{s}\'{a}kov\'{a} \citep{Ref16}, both consistent with VSMOW. A relatively high D/H ratio, three times VSMOW, was subsequently measured by \emph{Rosetta} in another Jupiter-family comet 67P/Churyumov-Gerasimenko \citep{Ref17}. The VSMOW D/H value measured in the Oort cloud comet C/2014 Q2 (Lovejoy, \citealt{Ref18}) suggests that the same isotopic diversity is present in the two comet families. 

The large variations in the D/H ratio in cometary water have been interpreted as reflecting their formation in different regions of the solar nebula. Models considering isotopic exchanges in an evolving accretion disk  predict an increase in the D/H ratio with increasing distance from the star \citep{Ref10}. The same isotopic diversity observed in both Oort cloud and Jupiter-family comets could then be explained by the recent evidence that the formation zones of the two families largely overlapped and extended over a broad range of heliocentric distances \citep{Ref19}.

In this Letter we present a new measurement of the D/H ratio in the Jupiter-family comet 46P/Wirtanen carried out using the Stratospheric Observatory for Infrared Astronomy (SOFIA). This comet, which was the initial target of the \emph{Rosetta} mission, has an orbital period of 5.439 yr and made a close approach to Earth (0.08 au) a few days after its perihelion passage on 2018 December 12 at 22:20 UT  (perihelion distance $q$ =  1.05536 au). Comet 46P/Wirtanen belongs to the category of hyperactive comets, emitting more water molecules than can be expected given the size of the nucleus, which is explained by the presence of sublimating water-ice-rich particles within the coma. Using a sample of comets with known D/H ratios in water and nucleus sizes, we show that a remarkable correlation is present between the D/H ratio and hyperactivity.

\section{SOFIA observations of comet 46P/Wirtanen}

Previous spectroscopic detections of HDO were obtained from observations of ro-vibrational and rotational transitions in the infrared and submillimeter domains \citep{Ref1}. Low-energy rotational transitions of water are not accessible from the ground or  suborbital platforms. However, the atmosphere at stratospheric altitudes is sufficiently transparent at the frequencies of water isotopologues. In particular, the 547 and 509 GHz $1_{1,0} - 1_{0,1}$ transitions of H$_2^{18}$O and HDO, previously observed in several comets by \emph{Herschel}, are now accessible from SOFIA and can be used to accurately measure the D/H isotopic ratio. This requires assumptions about the $^{16}$O/$^{18}$O isotopic ratio, which has been shown to be relatively uniform in comets, $500 \pm 50$, and close to the terrestrial ratio \citep{Ref1}. 

The close 2018 December apparition of comet 46P/Wirtanen provided an excellent opportunity to demonstrate the utility of SOFIA for D/H measurements. Observations presented here were carried out during five SOFIA flights between 2018 December 14 and 20 UT. During each flight, comet Wirtanen was observed in a single flight leg of about 3 hours (the longest time allowed by the flight planning). A typical observing sequence consisted of a 7 – 17 min on-source integration at the frequency of the $1_{1,0}-1_{0,1}$ transition of H$_2^{18}$O, followed by a  26 – 34 min on-source integration at the frequency of the $1_{1,0}-1_{0,1}$ transition of HDO. Monitoring the H$_2^{18}$O emission was important for averaging out possible variations in the water production rate during the period of the observations.  Additional observational details are provided in Appendix \ref{sec:obs}.

\begin{figure}
   \centering
   \includegraphics[width=0.85\columnwidth]{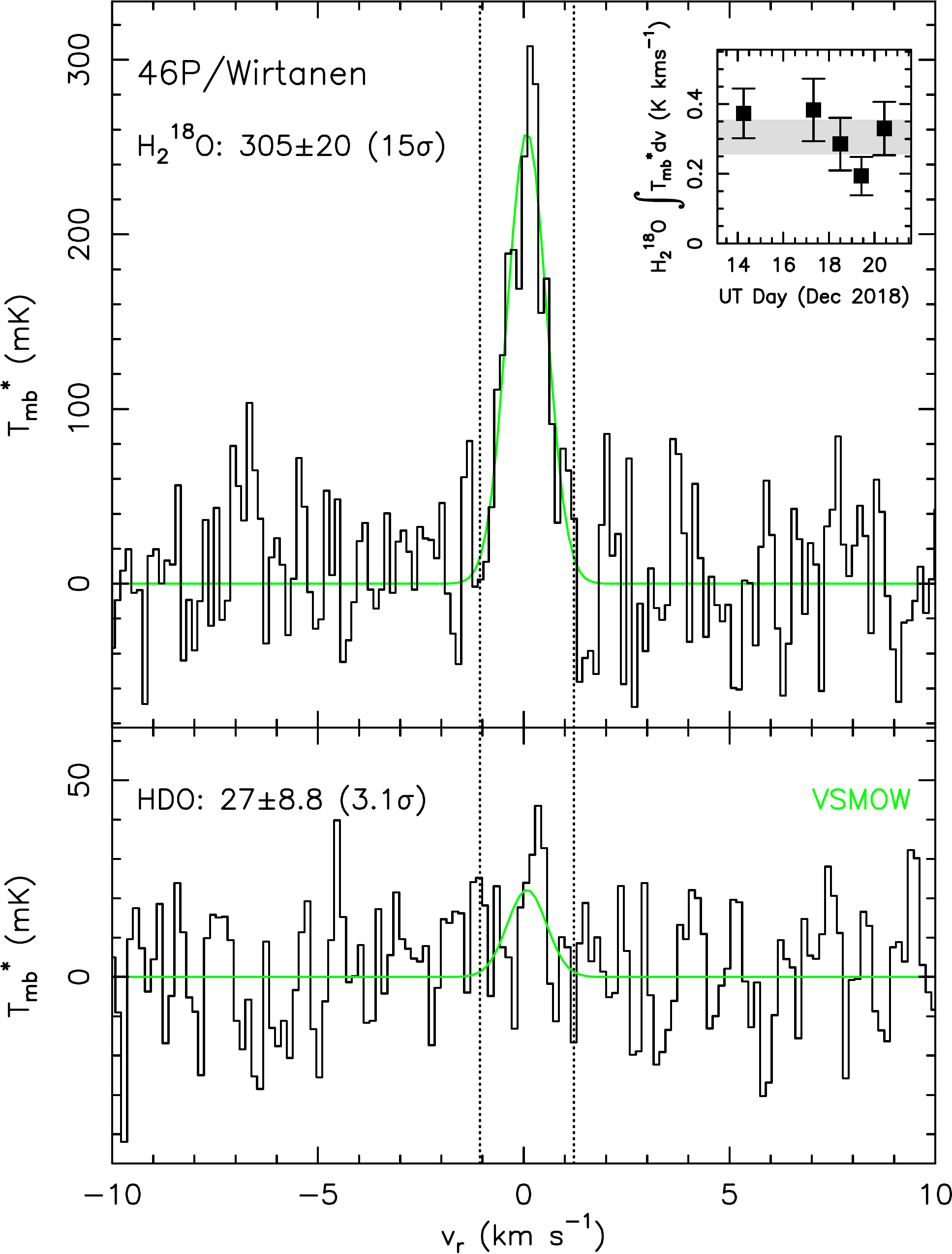}
   \caption{Spectra of the water isotopologues in comet 46P/Wirtanen. The $1_{1,0}-1_{0,1}$ H$_2^{18}$O and HDO transitions are shown in the upper and lower panels, respectively. The intensity scale is the main beam brightness temperature. The spectral resolution is 0.24 MHz, corresponding to approximately 0.14 km\,s$^{-1}$. A Gaussian fit to the H$_2^{18}$O spectrum (green line,  upper panel) gives a line center velocity $v_o=0.08 \pm 0.04$ km\,s$^{-1}$ and a full width at half maximum line width $\Delta v=1.09 \pm 0.09$ km\,s$^{-1}$. Vertical dotted lines indicate the velocity range used in computations of the integrated line intensities (–1.04 to 1.2 km\,s$^{-1}$). The green line in the lower panel shows the expected HDO line intensity assuming D/H equal to VSMOW. The inset in the upper panel shows the evolution of the H$_2^{18}$O integrated line intensity as a function of UT time. Error bars include statistical and calibration uncertainties, combined in quadrature, and the gray shaded area shows the corresponding uncertainty on the average H$_2^{18}$O line intensity (ensemble average).} 
         \label{fig1}
\end{figure}

Average spectra of the H$_2^{18}$O and HDO transitions are shown in Figure~\ref{fig1}.  The integrated H$_2^{18}$O line intensity is $305 \pm 20$ mK\,km\,s$^{-1}$ on the main beam brightness temperature scale (15.3 $\sigma$, average of all observations). The corresponding integrated line intensity of the HDO emission is $27 \pm 8.8$ mK\,km\,s$^{-1}$ (3.1 $\sigma$).  The resulting HDO/H$_2^{18}$O line intensity ratio is $0.089 \pm 0.034$,  compared to $0.094 \pm 0.009$ in comet 103P/Hartley \citep{Ref15}.  To model the water isotopologue emission, we used a cometary excitation model similar to that previously applied to \emph{Herschel} observations \citep{Ref15,Ref16}, and assumed a $^{16}$O/$^{18}$O isotopic ratio of 500 (see Appendix \ref{sec:model}). The resulting D/H ratio in water is $(1.61 \pm 0.65) \times 10^{-4}$, where the uncertainty includes statistical, calibration, modeling, and $^{16}$O/$^{18}$O isotopic ratio uncertainties, combined in quadrature. Comet 46P/Wirtanen is thus the third Jupiter-family comet with a D/H ratio consistent with the Earth’s ocean value.

\section{Correlation between the D/H ratio and hyperactivity}

When both the water production rate and the nucleus size are known, it is possible to compute the active fractional area of the nucleus (or active fraction) by dividing the active area by the total nucleus surface. Comets with high active fractions are referred to as hyperactive comets.  This hyperactivity requires an additional source of water vapor, explained by the presence of subliming icy grains in the coma that have been expelled from the nucleus. The archetype of a hyperactive comet is 103P/Hartley, studied by the Deep Impact spacecraft, for which both icy grains and water overproduction were observed \citep{Ref20,Ref36,Ref37}. Interestingly, the three Jupiter-family comets with a terrestrial D/H ratio, 46P, 103P, and 45P, all belong to the category of hyperactive comets. We therefore investigated quantitatively how the D/H ratio correlates with the active fraction using a sample of comets from the literature.

The active fraction was computed using a sublimation model and water production rates derived from Lyman-$\alpha$ observations by the Solar Wind Anisotropies (SWAN) instrument aboard the Solar and Heliocentric Observatory (SOHO) \citep{Ref29} (see Appendix \ref{sec:activefraction}). Since the SWAN field of view is large, water production rates include direct production from the nucleus surface and from subliming icy grains. We computed the active fraction using both production rates at 1 au and at  perihelion.

In the sample of comets with D/H determinations (or significant upper limits), only eight comets have a known nucleus size, most of them from spacecraft images or radar measurements (Appendix \ref{sec:activefraction}): 1P/Halley, 8P/Tuttle, 45P/Honda-Mrkos-Pajdu\v{s}\'{a}kov\'{a}, 46P/Wirtanen, 67P/Churyumov-Gerasimenko, 103P/Hartley, C/1996~B2 (Hyakutake), and C/1995~O1 (Hale-Bopp). We also consider the hyperactive comet C/2009~P1 (Garradd), whose nucleus effective radius has been constrained to be $<$ 5.6~km \citep{Ref39}. The effective nucleus radius of comet 46P/Wirtanen is estimated to 0.63 km from radar imaging.\footnote{https://uanews.arizona.edu/story/ua-researcher-captures-rare-radar-images-comet-46pwirtanen}

Figure 2 shows a striking anti-correlation between the D/H ratio and the active fraction computed at perihelion. The same trend for a D/H ratio decreasing towards the telluric value with increasing active fraction is observed when using the active fraction at 1 au from the Sun (Fig. A.1). Values for the D/H ratios are taken from the review of \citet{Ref1}, except for comet C/1996B2 (Hyakutake), for which we use a revised value of (1.85 $\pm$ 0.6) $\times$ 10$^{-4}$ (Appendix \ref{sec:hyakutake}). This long-period comet displayed outbursts and fragmentation events over a few months before and after perihelion, when it released icy grains and chunks, hence the large active fraction \citep[Fig. 2,][]{Combi2005}. The D/H ratio reported by \citet{Ref50} of (2.9 $\pm$ 1.0) $\times$ 10$^{-4}$ was measured during an outburst, with a large uncertainty mainly related to the scatter in reported water production rates. For this new evaluation, we used updated $Q$(H$_2$O) values \citep{Combi2005}. 

\begin{figure}
   \centering
   \includegraphics[width=0.99\columnwidth]{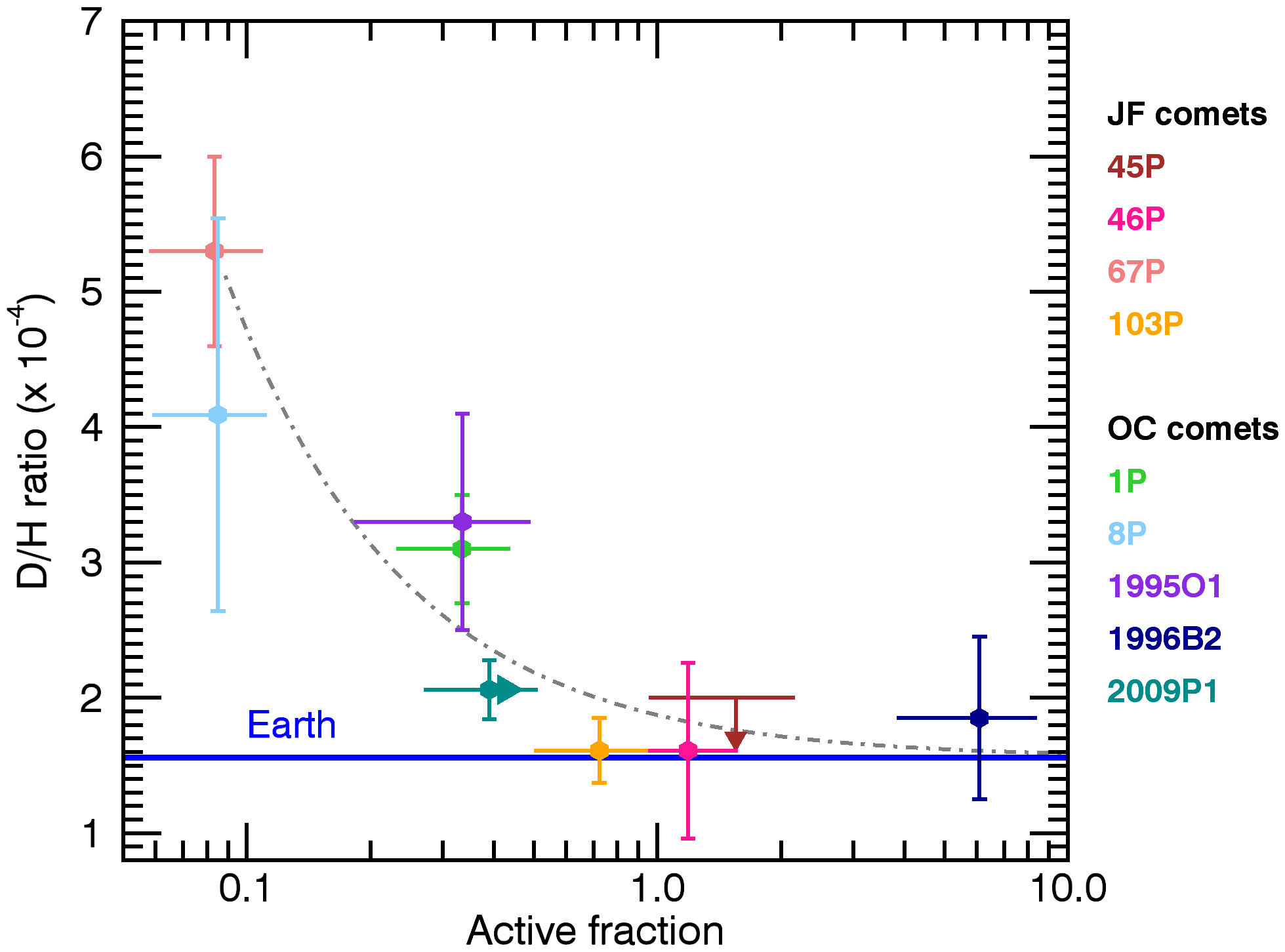}
   \caption{ D/H ratio in cometary water as a function of the active fraction computed from the water production rates measured at  perihelion. The uncertainties on the active fraction (horizontal error bars) include a 30\% uncertainty on the water production rates \citep{Ref29}  and the uncertainty on the nucleus size. The color of each symbol indicates a  comet; see legend at right, where  the
dynamical class is also indicated: Oort cloud (OC) or short-period Jupiter-family (JF) comets. The blue horizontal line corresponds to the VSMOW D/H value. The upper limit for the D/H ratio in comet 45P is indicated by a downward arrow and the lower limit for the active fraction in comet 2009P1 by a right arrow. The dash-dotted line shows the expected D/H assuming two sources of water: D-rich (3.5$\times$VSMOW) from the nucleus and D-poor (VSMOW). Comets with an active fraction equal to 0.08 are assumed to release only D-rich water. }
         \label{fig2}
\end{figure}

We  investigated the processes responsible for the excess of icy grains in hyperactive comets by considering a sample of 18 comets with determined nucleus sizes and water production rates at  perihelion (Appendix \ref{sec:activefraction}). As shown in Fig. 3, hyperactivity is not observed for comets with effective nucleus radii larger than 1.2~km (12 comets in our sample), whereas comets with smaller nuclei, though underrepresented considering the size distribution of cometary nuclei \citep{Fernandez2013}, are all hyperactive. This suggests that the large amount of subliming icy aggregates or chunks in hyperactive comets is not related to a higher ice/refractory content. A comparison between the well-studied comets 67P and 103P shows that even though the nucleus gas production is much lower in 103P than in 67P, owing to a smaller nucleus size (Fig. 3), the mass loss rate in chunks is larger for 103P \citep{Ref23}, thereby explaining its hyperactivity.  Estimates of the refractory-to-ice mass ratio in 67P \citep{Herique2016,Ref22,Ref23} converge to values between $\delta = 3$ and 7, matching the rough estimate of $\delta = 3$ for 103P \citep{Ref23}. 

Dust aggregates can be ejected by the sublimation of water ice if the gas pressure overcomes the tensile strength and gravitational pressure of the covering dust layer, which depend on the local gravity, hence nucleus size \citep{Ref21}. However, it remains to be demonstrated why small nuclei  eject chunks so efficiently.

\begin{figure}
   \centering
   \includegraphics[width=0.99\columnwidth]{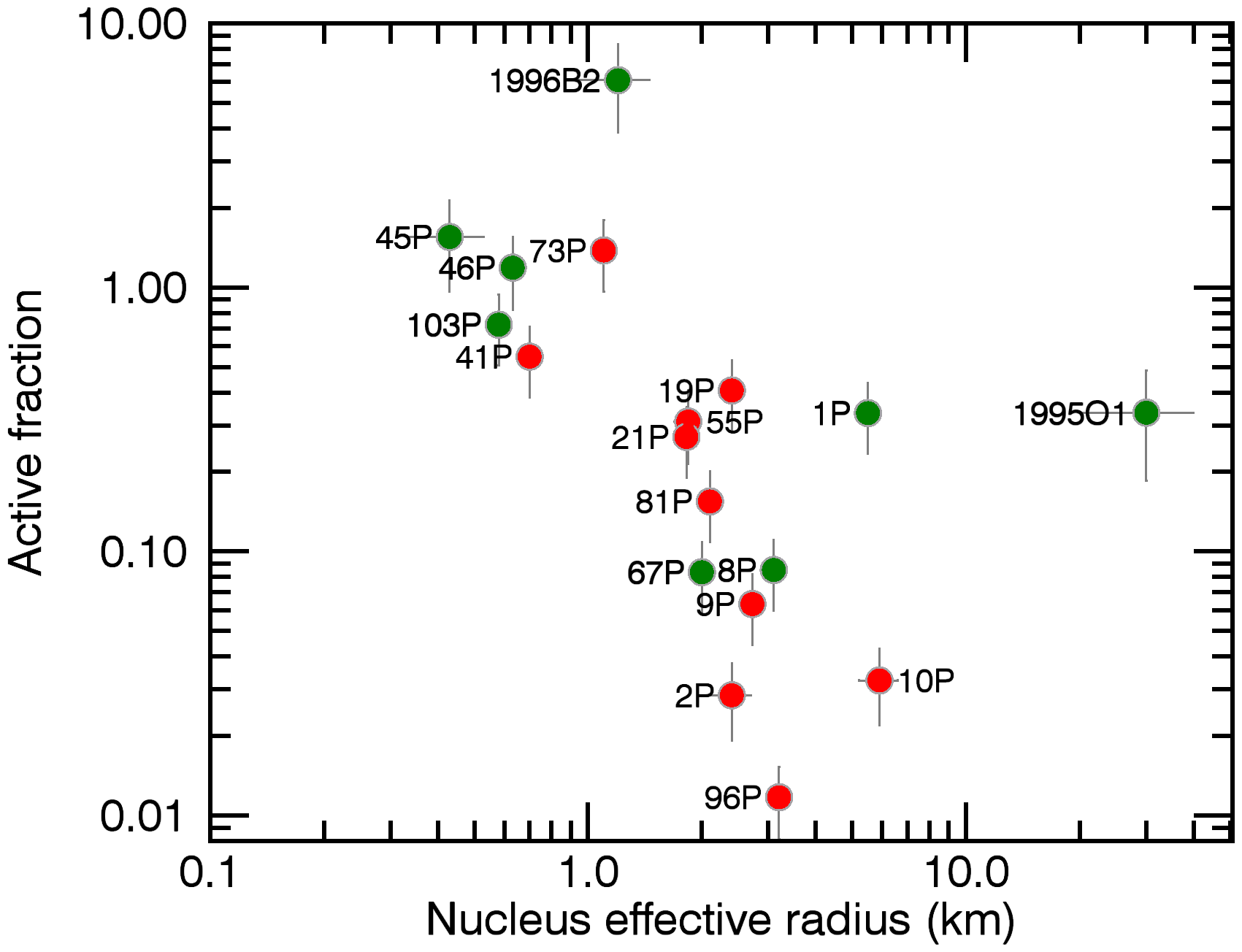}
   \caption{Active fraction at perihelion as a function of the nucleus size for a sample of 18 comets. The uncertainties in the active fraction (vertical error bars) include a 30\% uncertainty on the water production rates \citep{Ref29} and the uncertainty on the nucleus size.  The horizontal error bars show the uncertainties in the nucleus size. Green symbols refer to comets for which the D/H ratio in water has been measured. }
         \label{fig3}
\end{figure}

\section{Discussion}

Under the hypothesis that hyperactive comets belong to a population of ice-rich comets, their Earth-like D/H ratio would be consistent with their formation in the protoplanetary disk just outside the snow line where a large enhancement in the ice surface density is expected, thus favoring planetesimal formation \citep{Ref24}. This would explain the surprising result from \emph{Rosetta} that ice-poor, D-rich comets, such as comet 67P, are less rich in water than material from carbonaceous meteorites formed closer to the Sun \citep{Ref23}.  Alternatively, these hyperactive comets could have formed in the outermost regions of the solar nebula.  Indeed, modeling shows that a non-monotonic dependence of the D/H ratio in the solar nebula may be expected. The ratio would decrease again in the outer regions of the disk, because water molecules that underwent isotopic exchanges at high temperatures near the young star would have been transported outward during the early disk evolution \citep{Yang}.  The anti-correlation between hyperactivity and nucleus size appears inconsistent with the first explanation as planetesimals near the snow line are expected to undergo rapid growth.

An alternative explanation is that the isotopic properties of water outgassed from the nucleus surface and icy grains may be different, owing to fractionation processes during the sublimation of water ice.
The observed anti-correlation can  be reproduced with two sources of water contributing to the measured water production rate and the active fraction: D-rich water molecules released from the nucleus and an additional source of D-poor water molecules from sublimating icy grains (see dash-dotted line in Fig. 2).  Laboratory experiments on samples of pure ice show small deuterium fractionation effects \citep{Ref25}. In experiments with water ice mixed with dust, the released water vapor is depleted in deuterium, explained by preferential adsorption of HDO on dust grains \citep{Ref26}.  This effect goes in the opposite direction to the observed trend, while conversely the VSMOW D/H value from subliming icy grains is likely representative of buried nucleus water ice. Alternatively, a non-steady-state regime of water ice sublimation could explain the factor $2 - 3$ deuterium enrichment in water vapor released from the nucleus with respect to VSMOW \citep{Ref27}. However, this would occur in episodic time intervals along the comet orbit \citep{Ref27}, whereas the HDO/H$_2$O values measured for comet 67P at different periods are very uniform \citep{Ref28}. 

\section{Conclusion}

Enlarging the number of accurate D/H measurements in both Jupiter-family and Oort cloud comets is required to better constrain the observed correlation. Taking these measurements from the ground  is challenging. Nearly simultaneous spectroscopic observations of low-energy rotational lines of H$_2^{18}$O and HDO in a matching field of view encompassing a large fraction of the coma using the GREAT spectrometer aboard SOFIA can play a key role in this endeavor. In this context, the next close apparition of comet 67P/Churyumov-Gerasimenko in 2021 November will offer an excellent opportunity to re-measure the D/H ratio in this comet using spectroscopic techniques.

The understanding of the observed correlation calls for detailed investigations of the mechanisms leading to dust and chunk ejection, and cometary hyperactivity. Further experimental and modeling work on evaporative fractionation is also clearly needed, and may ultimately establish that all comets share the same Earth-like water D/H ratio, with profound implications on the early solar system and the origin of the Earth’s oceans.

\begin{acknowledgements}
Based on observations made with the NASA/DLR Stratospheric Observatory for Infrared Astronomy (SOFIA). SOFIA is jointly operated by the Universities Space Research Association, Inc. (USRA), under NASA contract NAS2-97001, and the Deutsches SOFIA Institut (DSI) under DLR contract 50 OK 0901 to the University of Stuttgart. GREAT is a development by the MPI für Radioastronomie and the KOSMA/Universität zu Köln, in cooperation with the DLR Institut für Optische Sensorsysteme, financed by the participating institutes, by the German Aerospace Center (DLR) under grants 50 OK 1102, 1103, and 1104, and within the Collaborative Research Centre 956, funded by the Deutsche Forschungsgemeinschaft (DFG). Mixers for Channel 1 of the 4GREAT (4G-1) instrument have been designed and developed by LERMA (Observatoire de Paris, CNRS, Sorbonne Université, Université de Cergy-Pontoise) in the framework of the \emph{Herschel}/HIFI project, with funding from the CNES. Part of this research was carried out at the Jet Propulsion Laboratory, California Institute of Technology, under a contract with the National Aeronautics and Space Administration. We thank the SOFIA project office for their excellent support and for adapting the operations and engineering support to the visibility constraints of the comet. We thank J. Blum, M. Fulle, and A. Morbidelli for the useful discussions. 
\end{acknowledgements}

\noindent \emph{Note added in proof:} After the present manuscript was accepted for publication, it was brought to our attention that the correlation between the active fraction and the nucleus size was independently found by Sosa \& Fern\'{a}ndez (2011, MNRAS 416, 767).

\begin{appendix}
\section{Appendix}

\subsection{Observations}
\label{sec:obs}
The observations of comet 46P/Wirtanen reported here were carried out using the GREAT heterodyne spectrometer \citep{Ref30} aboard SOFIA during five flights between 2018 December 14 and 20 UT, out of Palmdale, CA, USA. The instrument was operated in its upGREAT HFA/4GREAT (HFA/4G) 
configuration \citep{Ref31,Ref32}, which allows simultaneous observations at five different frequencies. The lowest frequency band, 4G-1, was used for the observations reported here. The tuning setup and the basic instrument characteristics are summarized in Table A.1. Although several other transitions of interest were covered in the higher-frequency channels, only OH was detected at a low signal-to-noise ratio, and the other upper limits are not constraining owing to the much higher system temperatures at these frequencies. 

\begin{table*}
\begin{center}  
\label{tab:inst}
\caption{Instrument tuning and performance.}
\begin{tabular}{cccccc}
\hline \hline
\rule[-3mm]{0mm}{8mm}Transition & $\nu$ & IF$_{\rm c}$ & T$_{\rm sys}$  & $\Theta$ & $\eta_{mb}$ \\
&  (GHz)      & (GHz)    &              (K)     & ($^{\prime\prime}$)  \\
\hline
H$_2^{18}$O ($1_{1,0}-1_{0,1}$) & 547.676440& 5.4 U & 317 & 50.3 & 0.63 \\
HDO ($1_{1,0}-1_{0,1}$) & 509.292420 & 6.2/5.45 U & 290 & 54.1 &0.63  \\
\hline
\end{tabular}
\end{center}

Note: Entries in the table are molecular transition, rest frequency (from JPL Molecular Spectroscopy Database; \citealt{Ref35}), intermediate frequency and sideband (carefully chosen by balancing between the atmospheric transmission in the upper and lower sidebands, the receiver performance, and possible line contamination from the image sideband), single sideband system temperature (average across the intermediate frequency band), FWHM beam width, and main beam efficiency. 
\end{table*}

\begin{table*}
\begin{center}  
\label{tab:obslog}
\caption{SOFIA observations of comet 46P/Wirtanen.}
\begin{tabular}{cccccccc}
\hline \hline
\rule[-3mm]{0mm}{8mm}Flight & UT Time & $r_h$ & $\Delta$  & $t$(H$_2^{18}$O) &  $\sigma$(H$_2^{18}$O) & $t$(HDO) &  $\sigma$(HDO) \\
&  (hr)      & (au)    &   (au)     & (min)  & (mK) & (min) & (mK) \\
\hline
1       & Dec 14, 4.89-7.47     &1.056  &0.079  &16.5   &80             &29.2   &43\\
2       &Dec 17, 7.56-9.68      &1.057  &0.078  &7.2    &125    &30.8   &38\\
3       & Dec 18, 9.59-12.17    &1.058  &0.078  &13.8   &112    &30.3   &37\\
4       & Dec 19, 9.78-12.00    &1.059  &0.079  &14.9   &85            &25.6    &42\\
5       & Dec 20, 9.83-12.33    &1.060  &0.081  &11.6   &105    &34.1   &31\\

\hline
\end{tabular}
\end{center}

Entries in the table are flight number, UT range in hours, average heliocentric and geocentric distance of the comet (au) as given by the ephemeris, total on-source integration time (minutes), and the resulting rms noise level in a 0.14 km s$^{-1}$ velocity channel (mK) for H$_2^{18}$O and HDO. The comet reached  perihelion on December 12 at 22:20 UT and made the closest approach to the Earth on December 16 at 13:06 UT. At a 0.08 au geocentric distance, the field of view on the comet was about 3000 km.
\end{table*}

The instrument was operated in double-beam chopped mode, with a chopper throw of 200$^{\prime\prime}$, at a rate of 2.5~Hz. The comet was tracked using an ephemeris based on the orbital solution 181-11 generated using JPL Horizons\footnote{https://ssd.jpl.nasa.gov/horizons.cgi}. Pointing was established by the telescope operators directly on the optical core of the comet to an accuracy of 2 – 3$^{\prime\prime}$. The local oscillator frequency was updated every 4 minutes according to the ephemeris, introducing a maximum velocity tracking error of about 0.002 km\,s$^{-1}$. Prior to the flight series, the optical axis of the GREAT instrument had been aligned to the optical imagers by observations of Mars. The main beam coupling efficiencies, also determined from observations of Mars, and the diffraction limited half-power beam widths are listed in Table A.1. The observations were performed at flight altitudes between 40,000 and 43,000 feet. Atmospheric conditions were typical for late autumn flights out of Palmdale, CA, with a residual water vapor column of about 15--20 $\mu$m, which resulted in typical single-sideband system temperatures $T_{sys}$ of about 300 K (Table A.1). 

The calibration at the frequencies of the HDO and H$_2^{18}$O lines is challenging. Locally, the lines are affected by the proximity of their rather narrow telluric counterparts, shifted by about $\pm 3$ km\,s$^{-1}$ at the time of observation. Moreover, the transmission at the H$_2^{18}$O frequency is strongly affected by the broad absorption of the nearby telluric H$_2^{16}$O line at 557 GHz (50\% transmission only). The spectra were corrected for atmospheric losses following the usual calibration scheme \citep{Ref34} based on two load signals (one at ambient temperature and one at a cold temperature) to determine the instrument gain and a blank sky signal (chopper off phase), to which the atmospheric model was fit in order to correct the observed signal to outside the atmosphere. The resulting calibration uncertainties at the frequencies of 509 and 547 GHz are 10\% and 15\%, respectively.

Calibrated spectra provided by the instrument pipeline were further reduced and analyzed using the IRAM Gildas software\footnote{http://www.iram.fr/IRAMFR/GILDAS}. A linear baseline was first subtracted from each scan and the resulting spectra were then averaged with 1/$\sigma^2$ weighting by radiometric noise. The observing log is shown in Table A.2. The heliocentric and geocentric distance of the comet changed only slightly during the period of the observations, with average values of 1.058 and 0.079 au, respectively. The total on-source integration time is 64 and 150 min for  H$_2^{18}$O and HDO, respectively. There is some evidence for day-to-day variations in the  H$_2^{18}$O line intensity (the intensity on December 19 UT seems lower compared to the other days; see inset in the upper panel of Fig. 1). However, the signal-to-noise ratio in the spectra obtained on the individual flights is limited. Since both HDO and  H$_2^{18}$O were observed on each flight and we do not expect day-to-day variations in the isotopic ratio, we use the average spectra in the subsequent analysis. 

\subsection{Modeling}
\label{sec:model}

To convert the observed line intensities into molecular production rates, we used an excitation model similar to that used in our earlier \emph{Herschel} studies \citep{Ref15,Ref16}. We  computed several models with different assumptions for the collisions with electrons. We used electron density factors $x_{ne} = 0$, 0.1, and 0.2 and a contact surface scaling factor $X_{re} = 0.5$ \citep{Ref35}. We used a variable temperature model with a temperature of 80 K out to a distance of 270~km (corresponding to approximately 10\% of the field of view), followed by a linear decrease to 12 K at a distance of 630 km, and a constant temperature at larger distances. A similar temperature profile provided a good fit to the observed spatial distribution of multiple water emission lines observed in comet 103P/Hartley. We also computed models with constant temperatures of 40 and 60 K, consistent with ground-based methanol observations carried out by members of our team using the IRAM 30m telescope. The maximum difference in the isotopic ratios retrieved using the various models is 15\%. In our analysis we used the average molecular production rates provided by the various models with a conservative modeling uncertainty of 10\%.

The observed line intensities lead to average HDO and H$_2^{18}$O production rates of $(2.5 \pm 0.9) \times 10^{24}$ s$^{-1}$ and $(1.5 \pm 0.3) \times 10^{25}$ s$^{-1}$, respectively, where the uncertainties include the statistical and calibration uncertainties, and a 10\% modeling uncertainty, combined in quadrature. Assuming a $^{16}$O/$^{18}$O isotopic ratio of $500 \pm 50$, we derive a H$_2^{16}$O production rate of $(7.7 \pm 1.5) \times 10^{27}$ s$^{-1}$. The resulting D/H ratio in water, $(1.61 \pm 0.65) \times 10^{-4}$, is close to the Earth’s ocean value. The uncertainty includes a 10\% uncertainty for the $^{16}$O/$^{18}$O isotopic ratio, combined in quadrature with the statistical, calibration, and modeling uncertainties.


\subsection{Computations of the active fraction}
\label{sec:activefraction}

To compute the active fractional area, we used the sublimation model of \citet{Ref38} for a rotational pole pointed at the Sun, which is identical to both the non-rotating case and to the case of zero thermal inertia.
This model is appropriate, as cometary nuclei have low thermal inertia \citep{Gulkis2015}.
We use tabulated values  for the average water sublimation rate per surface unit, $Z$, as a function of the heliocentric distance, $r_h$. Calculations are carried out for a Bond albedo of 0.05 and 100\% infrared emissivity. At $r_h$ = 1 au, $Z$ = $3.6 \times 10^{21}$ mol\,s$^{-1}$\,m$^{-2}$. The active area ($AA$) is obtained by dividing the water production rate by $Z$, and the active fractional area is obtained by dividing $AA$ by the nucleus surface area ($4 \pi r_ N^2$, where $r_N$ is the effective nucleus radius). We note that the derived active areas only provide  a crude estimation of the ice exposed to the solar radiation, because

the utilized sublimation model is simplistic. For example, the active fractions derived here differ by a large but constant factor from those computed assuming rapidly rotating nuclei \citep{Ref29}.

We consider water production rates $Q$(H$_2$O) derived from Lyman-$\alpha$ observations by the Solar Wind Anisotropies (SWAN) instrument aboard the Solar and Heliocentric Observatory (SOHO) \citep{Ref29}. Over 90\% of the observed hydrogen atoms are produced by H$_2$O or its photodissociation product OH. We use the reported absolute water production rates at $r_h=1$ au, and pre-/post-perihelion power laws with $r_h$ to derive water production rates at perihelion by averaging the production rates deduced from pre- and post-perihelion laws.   For some short-period comets, the SWAN survey includes multiple apparitions (e.g., 1997 and 2002 for 46P). In this case, we used the average results for multiple apparitions (Table 4 of \citealt{Ref29}). The SWAN survey does not include comet 1P/Halley, for which we assumed Q(H$_2$O) = $5 \times 10^{29}$ s$^{-1}$  ($r_h = 1.0$ au, with a $r_h^{-2}$ variation for Q(H$_2$O); \citealt{Ref40}). Water production rates used to compute the active fractions are listed in Table \ref{tab:results} (values at perihelion). For consistency, we did not consider the water production rate of 46P derived from the SOFIA 2018 observations (Sect.~\ref{sec:obs}), which is about a factor of two lower than the SWAN value (Table \ref{tab:results}), possibly because of the smaller projected field of view for this close apparition. This trend between aperture size and water production is observed for hyperactive comets.

To study how the active fraction correlates with the nucleus size, we added to our sample ten short-period comets with  well-characterized water production rates and nucleus sizes: 2P/Encke, 9P/Tempel 1, 10P/Tempel 2, 19P/Borrelly, 21P/Giacobini-Zinner, 41P/Tuttle-Giacobini-Kresak, 55P/Tempel-Tuttle, 73P/Schwassmann-Wachmann 3, 81P/Wild 2, and 96P/Machholz 1. Water production rates are from \cite{Ref29}, except for 9P, 10P, and 81P, for which we used measurements from \cite{Ref41}, \cite{Ref42}, and \cite{Ref43}.
Most of the nucleus sizes are given in \cite{Ref29}. For C/1995O1 (Hale-Bopp), we used a nucleus radius of 30~km \citep{Lamy2004}. For C/1996 B2 (Hyakutake), we used a value of 1.2~km \citep{Ref45}.  For 9P, 10P, and 81P the effective nucleus radii are 2.72, 5.9~km, and 2.1~km, respectively \citep{Ref46,Ref47,Ref48}. Comet properties and derived active fractions at perihelion are summarized in Table \ref{tab:results}.

There are strong biases in this study. Comets with low water production rates (below 5 $\times$ $10^{27}$ s$^{-1}$ at 1 au) are not considered.  In addition, small nuclei are largely underrepresented, considering estimates of the size distribution of short-period comets \citep{Fernandez2013}.  Therefore, our sample does not include small comets with  low active fractions, which may be present in the population of short-period comets, because their surface has been heavily mantled by refractory dust. However, large ($r_{\rm N} > 1$ km) comets are well represented in our sample. 

Figure A.2 shows the active fraction computed at perihelion plotted as a function of perihelion distance. There is no significant correlation between these two quantities. However, the two short-period comets 2P and 96P with small perihelion distances (0.34 and 0.12 au, respectively) have a low active fraction, which may be related to a gradual decrease in the ice-to-refractory ratio in subsurface layers \citep{Ref29}.
We  also looked for a possible dependence of the D/H ratio on the nucleus size, but do not find a statistically significant correlation between these two quantities.

\begin{table*}
\begin{center}  
\label{tab:results}
\caption{Comet properties and derived active fractions.}
\begin{tabular}{lcccccl}
\hline \hline
Comet & $q^a$ & $r_N^b$ & D/H$^c$ & $Q$(H$_2$O)$^d$ & Active fraction$^e$ \\
&(au) & (km) & & s$^{-1}$ & \\
\hline 
         1P &  0.59 &  5.50 & (3.10 $\pm$ 0.40) $\times$ 10$^{-4}$  & 1.5 $\times$ 10$^{30}$ & (3.3 $\pm$  1.0) $\times$ 10$^{-1}$ \\
        8P &  1.03 &  3.10 & (4.09 $\pm$ 1.45) $\times$ 10$^{-4}$  & 0.3 $\times$ 10$^{29}$ & (8.5 $\pm$  2.6) $\times$ 10$^{-2}$ \\
       45P &  0.53 &  0.43 $\pm$  0.10 & $<$ 2.00$\times$ 10$^{-4}$  & 0.5 $\times$ 10$^{29}$ & (1.6 $\pm$  0.6) $\times$ 10$^{0\phantom{0}}$ \\
       46P &  1.05 &  0.63 $\pm$  0.04 & (1.61 $\pm$ 0.65) $\times$ 10$^{-4}$  & 1.9 $\times$ 10$^{28}$ & (1.2 $\pm$  0.4) $\times$ 10$^{0\phantom{0}}$ \\
       67P &  1.24 &  2.00 & (5.30 $\pm$ 0.70) $\times$ 10$^{-4}$  & 0.9 $\times$ 10$^{28}$ & (8.3 $\pm$  2.5) $\times$ 10$^{-2}$ \\
      103P &  1.06 &  0.58 & (1.61 $\pm$ 0.24) $\times$ 10$^{-4}$  & 1.0 $\times$ 10$^{28}$ & (7.2 $\pm$  2.2) $\times$ 10$^{-1}$ \\
    C/1995O1 &  0.91 & 30.00 $\pm$ 10.00 & (3.30 $\pm$ 0.80) $\times$ 10$^{-4}$  & 1.7 $\times$ 10$^{31}$ & (3.4 $\pm$  1.5) $\times$ 10$^{-1}$ \\
    C/1996B2 &  0.23 &  1.20 $\pm$  0.25 & (1.85 $\pm$ 0.60) $\times$ 10$^{-4}$  & 0.9 $\times$ 10$^{31}$ & (6.1 $\pm$  2.2) $\times$ 10$^{0\phantom{0}}$ \\
    C/2009P1 &  1.55 & $<$  5.60 & (2.06 $\pm$ 0.22) $\times$ 10$^{-4}$  & 1.9 $\times$ 10$^{29}$ & (3.9 $\pm$  1.2) $\times$ 10$^{-1}$ \\
        2P &  0.33 &  2.40 $\pm$  0.30 &  & 0.8 $\times$ 10$^{29}$  & (2.9 $\pm$  0.9) $\times$ 10$^{-2}$ \\
        9P &  1.51 &  2.72 &  & 0.8 $\times$ 10$^{28}$  & (6.3 $\pm$  1.9) $\times$ 10$^{-2}$ \\
       10P &  1.42 &  5.90 $\pm$  0.70 &  & 2.2 $\times$ 10$^{28}$  & (3.2 $\pm$  1.0) $\times$ 10$^{-2}$ \\
       19P &  1.36 &  2.40 &  & 0.5 $\times$ 10$^{29}$  & (4.1 $\pm$  1.2) $\times$ 10$^{-1}$ \\
       21P &  1.04 &  1.82 $\pm$  0.05 &  & 0.4 $\times$ 10$^{29}$  & (2.7 $\pm$  0.8) $\times$ 10$^{-1}$ \\
       41P &  1.05 &  0.70 &  & 1.1 $\times$ 10$^{28}$  & (5.5 $\pm$  1.6) $\times$ 10$^{-1}$ \\
       55P &  0.98 &  1.84 $\pm$  0.15 &  & 0.5 $\times$ 10$^{29}$  & (3.1 $\pm$  1.0) $\times$ 10$^{-1}$ \\
       73P &  0.94 &  1.10 $\pm$  0.03 &  & 0.9 $\times$ 10$^{29}$  & (1.4 $\pm$  0.4) $\times$ 10$^{0\phantom{0}}$ \\
       81P &  1.60 &  2.10 &  & 1.0 $\times$ 10$^{28}$  & (1.5 $\pm$  0.5) $\times$ 10$^{-1}$ \\
       96P &  0.12 &  3.20 $\pm$  0.20 &  & 0.4 $\times$ 10$^{30}$  & (1.2 $\pm$  0.4) $\times$ 10$^{-2}$ \\
\hline
\end{tabular}
\end{center}

Notes: (a) Perihelion distance.  (b) Nucleus effective radius. See references in \citet{Ref29} and Sect.~\ref{sec:activefraction}. For spacecraft and radar data, the error was assumed to be insignificant.  (c) D/H in water. References in the review of \citet{Ref1}, except for 67P \citep{Ref17}, C/1996B2 (this work), and 46P (this work).  (d) Water production rate at perihelion distance. From SWAN measurements \citep{Ref29}, except for 1P \citep{Ref40}, 9P \citep{Ref41}, 10P \citep{Ref42}, and 81P \citep{Ref43}. See Sect.~\ref{sec:activefraction}. (e) Active fraction at perihelion distance.

\end{table*}

\begin{figure}
   \centering
   \includegraphics[width=0.99\columnwidth]{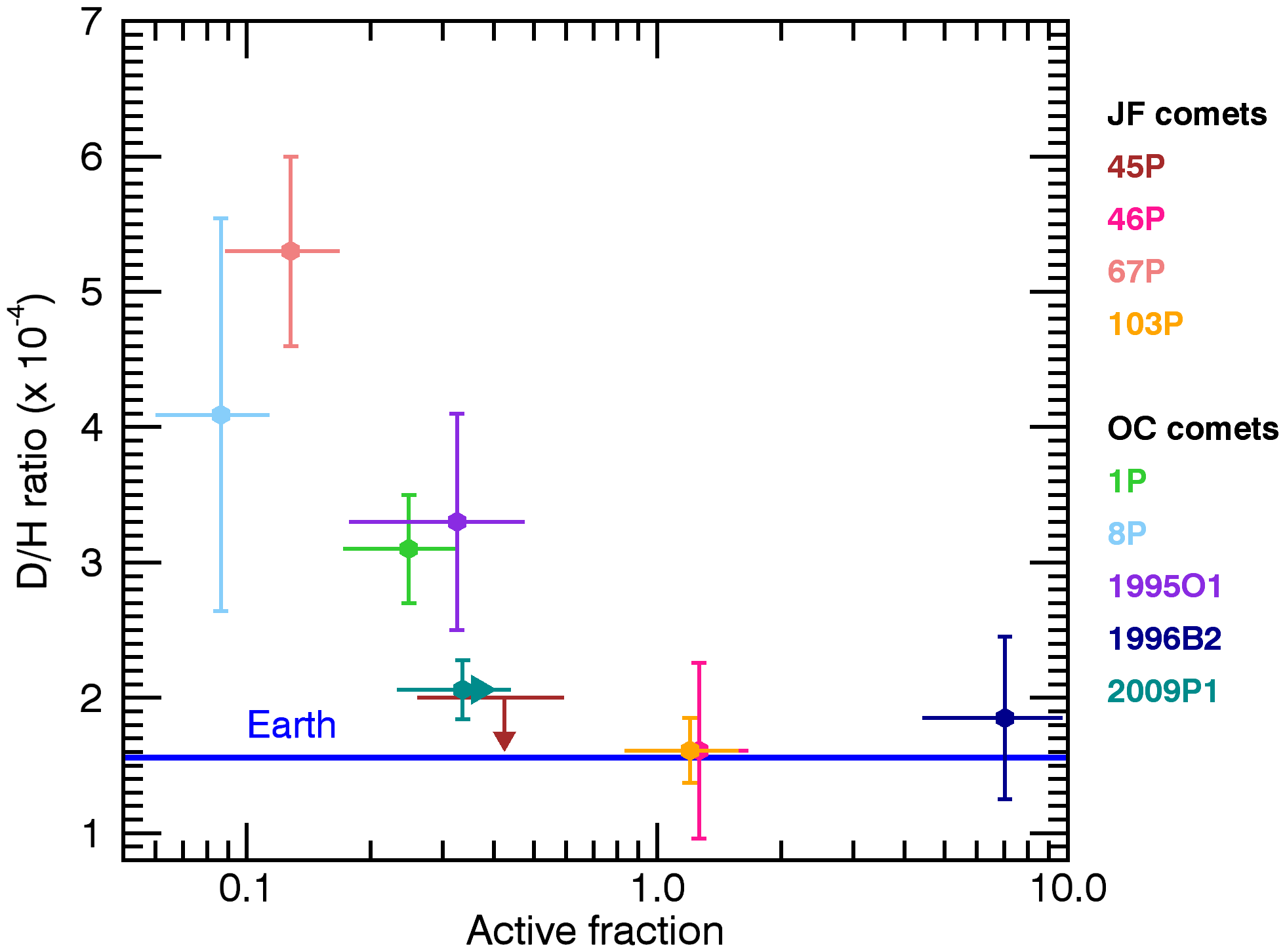}
   \caption{D/H ratio in cometary water as a function of the active fraction computed based on the water production rates at 1 au from the Sun.  The color of each symbol indicates a  comet; see legend at right, where
 the
dynamical class is also indicated: Oort cloud (OC) or short-period Jupiter-family
(JF) comets. The uncertainties on the active fraction (horizontal bars) include a 30\% uncertainty on the water production rates  \citep{Ref29} and the uncertainty on the nucleus size. }
         \label{figa1}
\end{figure}

\begin{figure}
   \centering
   \includegraphics[width=0.99\columnwidth]{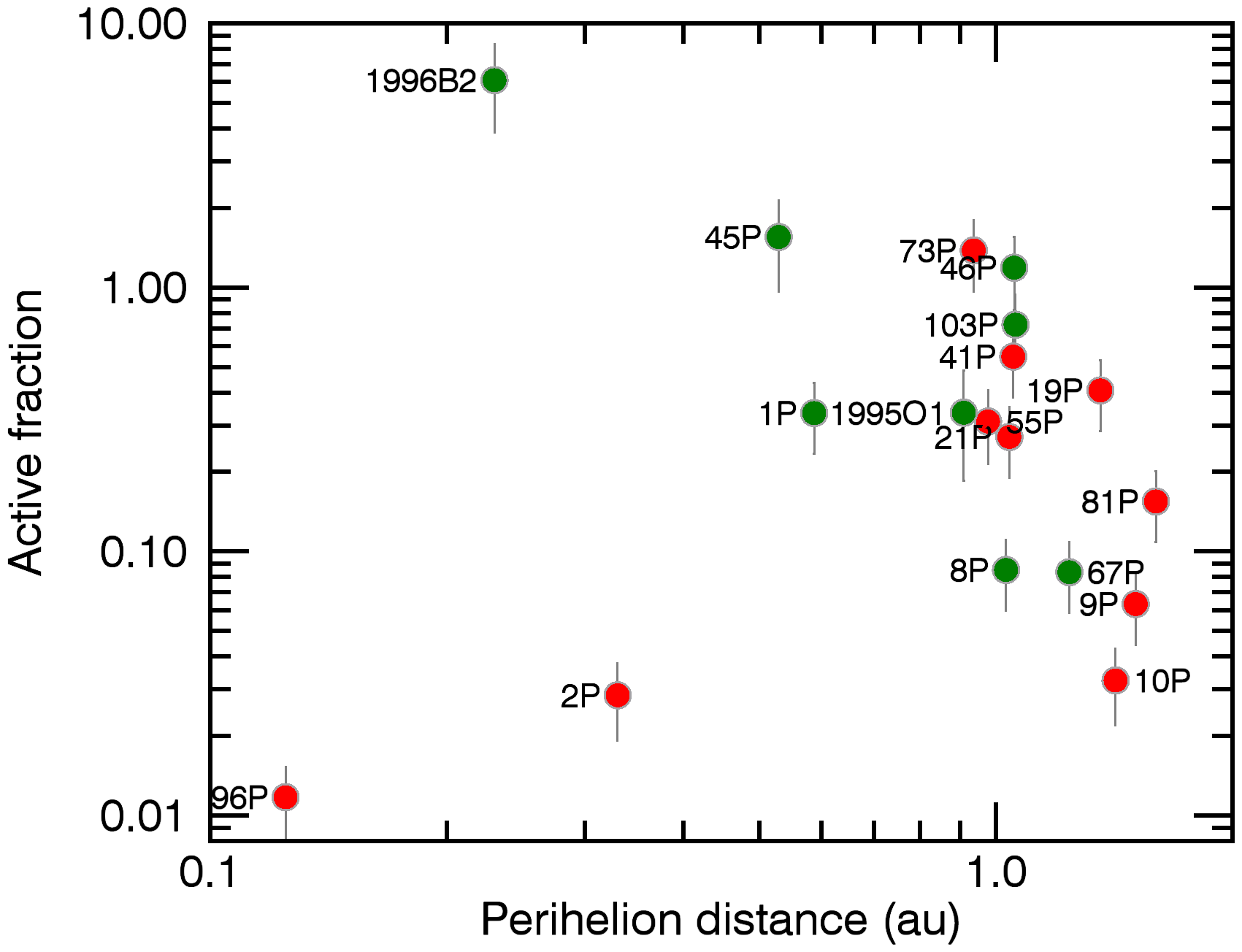}
   \caption{Active fraction at perihelion as a function of perihelion distance for a sample of 18 comets. Green symbols refer to comets for which the D/H ratio in water has been measured. The uncertainties on the active fraction (vertical bars) include a 30\% uncertainty on the water production rates \citep{Ref29} and  the uncertainty on the nucleus size. }
         \label{figa2}
\end{figure}

\subsection{D/H ratio in comet C/1996 B2 (Hyakutake)}
\label{sec:hyakutake}

The HDO production rate in comet Hyakutake was measured to be ($1.20 \pm 0.28$) $\times 10^{26}$ s$^{-1}$, averaging data obtained on March 23.5 and 24.5 UT, 1996 with the Caltech Submillimeter Observatory (CSO) \citep{Ref50}. A D/H ratio of ($2.9 \pm 1.0$) $\times 10^{-4}$ was derived using a water production rate of ($2.1 \pm 0.5$) $\times 10^{29}$ s$^{-1}$, corresponding to the average of reported measurements using observations of H$_2$O (IR), OH (UV, radio), H Lyman-$\alpha$, and OI (optical).  An updated analysis of the H Lyman-$\alpha$ SWAN observations \citep{Combi2005} indicates a higher water production rate than that adopted in \citet{Ref50}. \citet{Combi2005} compare SWAN retrievals to other $Q$(H$_2$O) measurements and conclude that there is a relatively good agreement, taking into account modeling induced differences. Using the daily tabulated $Q$(H$_2$O) values in Table~2 of \citet{Combi2005} and using interpolation, we derive $Q$(H$_2$O) = $3.23 \times 10^{29}$ s$^{-1}$ during the time of the HDO observations. Assuming a 25\% uncertainty for $Q$(H$_2$O) \citep{Combi2005}, the revised D/H ratio in comet Hyakutake is then ($1.85 \pm 0.6$) $\times 10^{-4}$.

HDO was detected in comet Hyakutake during an outburst. We therefore re-evaluated the HDO/H$_2$O production rate ratio using the methanol lines observed in the same spectrum as HDO and the re-evaluated CH$_3$OH average abundance relative to water outside the outburst period. Using water production rates from \citet{Combi2005} and the methanol production rates derived from JCMT, PdBi, and CSO observations before and after the 19--24 March outburst period \citep{Biver1999,Lis1997}, we find $Q$(CH$_3$OH)/Q(H$_2$O) = $0.013 \pm 0.003$. The computation of the HDO and methanol production rates from the CSO data of 23.5 and 24.5 March yields $Q$(HDO)/$Q$(CH$_3$OH) = $0.026 \pm 0.005$. Hence, we infer $Q$(HDO)/$Q$(H$_2$O = ($3.4 \pm 1$) $\times 10^{-4}$ and consequently D/H = ($1.7 \pm 0.5$)$\times 10^{-4}$, in agreement with the value derived above.

\subsection{D/H ratio in comet C/2014 Q2 (Lovejoy)}

A D/H ratio in water of $(1.4 \pm 0.4) \times 10^{-4}$ was measured in comet C/2014 Q2 (Lovejoy) based on the ground-based detection of a millimeter HDO line \citep{Ref18}. Instead, infrared observations undertaken with a much smaller field of view \citep{Ref51} yield D/H = $(3.02 \pm 0.87) \times 10^{-4}$. The inconsistency between the two values, which is marginal when considering the uncertainties of the two measurements, can be explained if this comet is a hyperactive comet. The VSMOW value measured in the millimeter would characterize water subliming from grains, whereas the value obtained in the IR would sample mainly water released directly from the nucleus. The size of the nucleus of this comet is currently unknown, so the active fraction cannot be computed. 

\end{appendix}

\begin{thebibliography}{}

\bibitem[A'Hearn et al. (1989)]{Ref46} A’Hearn, M. F., et al. 1989, \apj, 347, 1155
  
\bibitem[A'Hearn et al. (2005)]{Ref47} A’Hearn, M. F., et al. 2005, Science 310, 258
  
\bibitem[Alexander et al. (2012)]{Ref12} Alexander, C. M. O’D., et al. 2012, Science 337, 721

\bibitem[Altwegg et al. (2015)]{Ref17} Altwegg, K., et al. 2015, Science 347, 1261952

\bibitem[Altwegg et al. (2017)]{Ref28} Altwegg, K., et al. 2017, Phil. Trans. Royal Soc. Series A, 375, 20160253

\bibitem[Bischoff et al (2019)]{Ref21} Bischoff, D., et al. 2019, MNRAS 483, 1202

\bibitem[Biver et al.(1999)]{Biver1999} Biver, N., et al.\ 1999, \aj, 118, 1850 

\bibitem[Biver et al. (2007)]{Ref41} Biver, N., et al. 2007, Icarus 187, 253
  
\bibitem[Biver et al. (2012)]{Ref42} Biver, N., et al. 2012, A\&A, 539, A68

\bibitem[Biver et al. (2016)]{Ref18} Biver, N., et al. 2016, A\&A, 589, A78 

\bibitem[Bockel{\'e}e-Morvan et al.(1998)]{Ref50} Bockel{\'e}e-Morvan, D., et al.\ 1998, \icarus, 133, 147 

\bibitem[Bockel\'{e}e-Morvan et al. (2015)]{Ref1} Bockel\'{e}e-Morvan, D., et al. 2015, Space Sci. Rev. 197, 47
  
\bibitem[Boissier et al. (2013)]{Ref39} Boissier, J., et al. 2013, A\&A, 557
  
\bibitem[Brasser \& Morbidelli (2013)]{Ref19} Brasser, R., \& Morbidelli, A.  2013, Icarus 225, 40

\bibitem[Brownlee et al. (2004)]{Ref48} Brownlee, D. E., et al. 2004. Science 304, 1764

\bibitem[Ceccarelli et al. (2014)]{Ref7} Ceccarelli, C., et al.2014, in Protostars and Planets VI, p. 859

\bibitem[Combi et al.(2005)]{Combi2005} Combi, M.~R., et al.\ 2005, \icarus, 177, 228 
  
\bibitem[Combi et al. (2019)]{Ref29} Combi, M. R., et al. 2019, Icarus 317, 610

\bibitem[Cooke et al. (2014)]{Ref5} Cooke, R. J., et al. 2014, \apj, 781, 31
  
\bibitem[Cowan \& A'Hearn (1979)]{Ref38} Cowan, J. J., \& A’Hearn, M. F. 1979, Moon and Planets 21, 155
  
\bibitem[Deloule et al. (1998)]{Ref11} Deloule, E., et al. 1998, Geoch. Cosmoch. Acta 62, 3367

\bibitem[de Val-Borro et al. (2010)]{Ref43} de Val-Borro, M., et al. 2010, A\&A, 521, L50
  
\bibitem[Drouart et al. (1999)]{Ref10} Drouart, A., et al. 1999,Icarus 140, 129

\bibitem[Dur\'{a}n et al. (2017)]{Ref32} Dur\'{a}n, C. A., et al. 2017, ISST Proc. 27
  
\bibitem[Feldman et al. (1997)]{Ref40} Feldman, P. D., et al. 1997, \apj, 475, 829
  
\bibitem[Fern{\'a}ndez et al.(2013)]{Fernandez2013} Fern{\'a}ndez, Y.~R., et al.\ 2013, \icarus, 226, 1138 
  
\bibitem[Fulle et al. (2019)]{Ref23} Fulle, M., et al. 2019. MNRAS 482, 3326
  
\bibitem[Geiss \& Gloeckler (1998)]{Ref6} Geiss, J. \& Gloeckler, G. 1998, Sp. Sci. Rev. 84, 239

\bibitem[Guan et al. (2012)]{Ref33} Guan, X., et al. 2012, A\&A, 542, L4

\bibitem[Gulkis et al.(2015)]{Gulkis2015} Gulkis, S., et al.\ 2015, Science, 347, aaa0709
  
\bibitem[Hallis (2015)]{Ref14} Hallis, L. J. 2015, Science 350, 795

\bibitem[Harmon et al. (1997)]{Ref45} Harmon, J. K., et al. 1997, Science 278, 1921
 
\bibitem[Hartogh et al. (2011)]{Ref15} Hartogh, P., et al. 2011, Nature 478, 218

\bibitem[Herique et al.(2016)]{Herique2016} Herique, A., et al.\ 2016, \mnras, 462, S516 

\bibitem[Heyminck et al. (2012)]{Ref30} Heyminck, S., et al. 2012,  A\&A, 542, L1
  
\bibitem[Kelley et al. (2015)]{Ref36} Kelley, M. S. P., et al. 2015, Icarus 262, 187
  
\bibitem[Kelley et al. (2013)]{Ref37} Kelley, M. S. P., et al. 2013, Icarus 222, 634

\bibitem[Lamy et al.(2004)]{Lamy2004} Lamy, P.~L., et al.\ 2004, Comets II, 223 

\bibitem[L\'{e}cuyer et al. (2017)]{Ref25} L\'{e}cuyer, C., et al. 2017, Icarus 285, 1
    
\bibitem[Lis et al.(1997)]{Lis1997} Lis, D.~C., et al.\ 1997, \icarus, 130, 355 

\bibitem[Lis et al. (2002)]{Ref8} Lis, D. C., et al. 2002, \apj, 571, L55
  
\bibitem[Lis et al. (2013)]{Ref16} Lis, D. C., et al. 2013, \apj, 774, L3

\bibitem[O'Brien et al. (2018)]{Ref4} O’Brien, D. P., et al. 2018, Space Sci. Rev. 214, 47 

\bibitem[McCanta et al. (2008)]{Ref13} McCanta, M. C., et al. 2008, Geochem. Astrochim. Acta 72, 1401

\bibitem[Moores et al. (2012)]{Ref26} Moores, J. E., et al. 2012, Planetary Science 1, 2
    
\bibitem[Paganini et al. (2017)]{Ref51} Paganini, L., et al. 2017, \apj, 836, L25    
    
\bibitem[Parise et al. (2004)]{Ref9} Parise, B., et al. 2004, A\&A, 416, 159

\bibitem[P\"{a}tzold et al. (2019)]{Ref22} P\"{a}tzold, M., et al. 2019, MNRAS  483, 2337

\bibitem[Pickett et al. (1998)]{Ref34} Pickett, H. M., et al. 1998, J. Quant. Spectrosc. \& Rad. Transfer 60, 883
  
\bibitem[Podolak et al. (2002)]{Ref27} Podolak, M., et al. 2002, Icarus 160, 208
  
\bibitem[Protopapa et al. (2014)]{Ref20} Protopapa, S., et al. 2014, Icarus 238, 191
  
\bibitem[Risacher et al. (2018)]{Ref31} Risacher, C., et al. 2018, J. of Astron. Instr. 7, 1840014
  
\bibitem[Schoonenberg \& Ormel (2017)]{Ref24} Schoonenberg, D., \& Ormel, C.W. 2017, A\&A, 602, A21  
  
\bibitem[van Dishoeck et al. (2014)]{Ref3} van Dishoeck, E. F., et al. 2014, in Protostars and Planets VI, p. 835

\bibitem[Westall (2018)]{Ref2} Westall, F. 2018, Space Sci. Rev. 214, 50

\bibitem[Yang et al.(2013)]{Yang} Yang, L., et al..\ 2013, \icarus, 226, 256 
  
\bibitem[Zakharov et al. (2007)]{Ref35} Zakharov, V., et al. 2007, A\&A, 473, 303


\end{thebibliography}
\end{document}